\documentclass[letter,aps,twocolumn,showpacs,preprintnumbers,amsmath,amssymb]{revtex4}
\usepackage{amsmath,amssymb,graphicx}
\usepackage {amssymb}
\usepackage{multirow}
\usepackage{color}
\newcommand{\nc}{\newcommand}
\nc{\ba}{\begin{eqnarray}} \nc{\ea}{\end{eqnarray}}
\nc{\be}{\begin{equation}} \nc{\ee}{\end{equation}}

\newcommand\s{\sigma}

\newcommand\e{\epsilon}

\newcommand\la{\lambda}

\newcommand\te{\theta}

\nc{\ga}{\gamma} \nc{\x}{{\bf x }} \nc{\kk}{{\bf k }} \nc{\f}{{\bf f
}} \nc{\T}{ \theta (s_i (t)- \s) } \nc{\TT}{ \theta (s_i (t_{ r \, i
} )- \s) } \nc{\br}{   (s_i (t)- \s)  } \nc{\fa}{\phi_1}
\nc{\fb}{\phi_2}

\input{epsf.sty}
\begin{document}
\title{Baryon production from embedded metastable strings}
\preprint{MIT-CTP 4511}
\author{ Johanna Karouby$^1$ and Ajit Mohan Srivastava$^2$}

\affiliation{
$^1$  Center for Theoretical Physics and Department of Physics, \\ Massachusetts Institute of Technology, Cambridge, Massachussetts 02139, USA \\
$^2$ Institute of Physics, Sachivalaya Marg, Bhubaneswer 751005, India}

\begin{abstract}

We quantify the baryon anti-baryon production generated by a metastable cosmic 
string, similar to the embedded pion string. More precisely, we study skyrmion production mediated by instantons  
generated by a pion-like metastable string in contact with a thermal bath,
and interpret these Skyrmions as baryons. As shown in a previous work, the core of such a metastable string can melt due to quantum tunneling in the charged field direction. The specific configuration of our string containing 4 scalar fields out of equilibrium in contact with a thermal bath is shown to yield  skyrmion production with partial or integer winding number. In this work, we describe and quantify this skyrmion production per unit length of the string. We also 
evaluate the skyrmion-anti skyrmions production by a dense string network by 
invoking similarity with the Skyrmion production in a phase transition.

 \end{abstract}

\maketitle
\section{Introduction}

In this article we present a mechanism to generate baryons from an
embedded metastable cosmic string. To estimate the baryon production we compute the production of Skyrmions by the string (see e.g. \cite{ShellVil, HK, RHBrev} for overviews on cosmic strings). Skyrmions are global textures and are characterized by the non trivial third homotopy group of the vacuum manifold.
An important characteristic of Skyrmions is that their field configuration takes on the vev everywhere and thus skyrmions only have gradient energy \cite{Benson:1993at,Preskill:1992bf,Hindmarsh:1992yy}.This is for example not the case for strings which have a restored symmetry in their core.

Interestingly, skyrmions can also describe baryons in the context of the The Skyrme model : Solitons in the linear sigma model can be interpreted as the baryons of 
QCD\cite{Skyrme:1961vq,Skyrme:1962vh,bal,Adkins:1983ya}.
It is therefore possible to compute the baryons productions from a pion
string by computing the instanton production.

In the following we study the decay of the false vacuum into the true ground state for a small region of the metastable string and compute
the production of Skyrmions. We consider the process in which a segment of
the string decays into Skyrmion-anti-Skyrmion pairs. We calculate the
decay rate for this process in order to quantify the production of Skyrmions
that can then be interpreted as baryons. As a result, the metastable string, besides having standard cosmological impact (see for e.g \cite{Brandenberger:1994bx,CSmag,Tashiro:2012pp,Bevis:2007gh,Bhattacharjee:1998qc}) could
also contribute to baryon density fluctuations through skyrmion 
production by the string network.

 We consider a certain type of metastable string  and focus on field-theoretic strings coupled to a thermal bath of photons in models analogous to the Ginzburg-Landau theory of superconductivity \cite{Nielsen:1973cs}.
The key idea is to compute the Skyrmions production rate for metastable strings previously stabilized by a photon plasma \cite{Karouby:2012yz}\cite{Karouby:2012kz}. More precisely , we consider an embedded string made of two non-thermal complex scalar fields : one is neutral $\phi$ and the other one is charged $\pi_c$.
Since charged fields are naturally coupled to photon gauge fields, a thermal bath of photons can easily be implemented in such cases. We consider a temperature below a certain critical temperature where the scalar fields are still out-of-equilibrium. 

In our approach we focus on first order phase transitions.
Such transitions usually imply a period of non-equilibrium physics and typicallyoccur via bubble nucleation.  The result of threose bubbles nucleations is 
the melting of the string and, as we will show, possible formation of 
skyrmions when two instantons meet.

The outline of the paper is as follows. In section II, we introduce our 
embedded metastable string and the linear sigma model of QCD. We recall 
results of bubble nucleation in this context.
In section III, we describe the string decay through meson production.
In section IV, we describe the physical picture of the string decay 
by baryon (Skyrmion) production. In section V, we provide estimate of
the number of produced Skyrmions per unit length per Hubble time.
Here we also discuss the production of baryons by a dense string network
by noting its similarity with Skyrmion production in a phase transition.
Section VI presents conclusions.

\section{The metastable string}

Our toy model for the metastable string is similar to the pion string  well described by the linear sigma model in the low energy QCD regime.
Previously, some of us found the effective potential at temperature T for out of equilibrium fields in contact with a photon plasma. 
This plasma stabilizes the pion string which is topologically unstable,
and makes the string metastable for a certain range of parameters such that $ \frac{3 e^4}{8\pi^2} \leq \la \ll e^2 \label{lam1}$\cite{Karouby:2012kz}.
In this case the model is analogous to the Ginzburg-Landau theory of superconductivity.

\subsection{The pion string}

As a toy model for the analytical study of the stabilization of
embedded defects by plasma effects we consider the chiral limit of the
QCD linear sigma model, involving the sigma field $\sigma$ and the
 pion triplet ${\vec \pi} = (\pi^0, \pi^1, \pi^2)$, given by the
Lagrangian
\begin{equation}
\label{lag1}
{\cal L}_0 \, = \, {1 \over 2} \partial_{\mu} \sigma \partial^{\mu}
\sigma + {1 \over 2} \partial_{\mu} {\vec \pi} \partial^{\mu} {\vec
\pi} - {\lambda \over 4} (\sigma^2 + {\vec \pi}^2 - \eta^2)^2 \, ,
\end{equation}
where $\eta^2$ is the ground state expectation value of $\sigma^2 + {\vec
\pi}^2$. In the following, we denote the potential in
(\ref{lag1}) by $V_0$.

Two of the scalar fields, the $\sigma$ and $\pi_0$, are electrically
neutral, the other two are charged. Introducing the coupling to
electromagnetism (through covariant derivatives) it is convenient to write the scalar field sector
${\cal L}$ of the resulting Lagrangian in terms of the complex scalar
fields
\begin{equation}
\label{pi}
\pi^+ \, = {1 \over {\sqrt{2}}} (\pi^1 + i \pi^2) \, , \,\,\,
\pi^- \, = {1 \over {\sqrt{2}}} (\pi^1 - i \pi^2) \, 
\end{equation}
According to the minimal coupling prescription we obtain
\begin{equation}
\label{lag2}
{\cal L} \, = \, {1 \over 2} \partial_{\mu} \sigma \partial^{\mu}
\sigma + {1 \over 2} \partial_{\mu} \pi^0 \partial^{\mu} \pi^0 +
D_{\mu}^+ \pi^+ D^{\mu -} \pi^- - V_0 \, 
\end{equation}
where
$D_{\mu}^+ \, = \, \partial_{\mu} + i e A_{\mu} \, , \,\,\,\, D_{\mu}^-
\, = \, \partial_{\mu} - i e A_{\mu} \, $.

Effective pion-photon interactions appear through the covariant derivative in (\ref{lag2}).
Suppose we have standard $S_3$ vacuum manifold. Then the first homotopy group is trivial
and thus there are no topologically stable string defects.
Note that if we restrict the field configuration to have
$\langle\pi_c\rangle \neq 0$, then the U(1) of electromagnetism 
gets broken and there is a magnetic flux of $\frac{2\pi}{e}$ 
in the core of the string.

Considering a specific configuration where $\sigma^2 +{\pi_0}^2 =\eta^2$
and  $\pi^+ = \pi^-=0$ the vacuum manifold $S^3$ reduces to $S^1$ and string configurations exist
but are not topologically stable since $\Pi_1(S^3)=1$. We have shown in
ref. \cite{Karouby:2012yz} that, using plasma effect and considering temperature below the confinement scale,
we have an embedded neutral string 
As usual, $\eta$ vanishes in the core of the string, but in addition the two charged fields vanish everywhere.
However this configuration is now metastable and the charged fields can take on non zero value only after quantum tunneling.

For simplicity we  work with 2 complex scalar fields, one charged that we call $\pi_c = \pi_1 +i\pi_2 $ and one uncharged,
$\phi = \s+i \pi_0$.

Considering a thermal bath and two non-equilibrium charged fields
gives rise to a new effective potential which at very  high-temperatures  $\frac{|\pi_c|}{T} \ll 1$  yields \cite{Karouby:2012yz}

\be
V_{eff}(\phi,\pi_c,T) \simeq \frac{\la}{4}(|\phi|^2+|\pi_c|^2-\eta^2)^2+\frac{e^2|\pi_c|^2 }{12}T^2-\frac{e^3|\pi_c|^3 }{6\pi}T
\label{veff}
\ee

\hspace{-0.5cm}
\begin{figure}[h]
\includegraphics[scale=0.20]{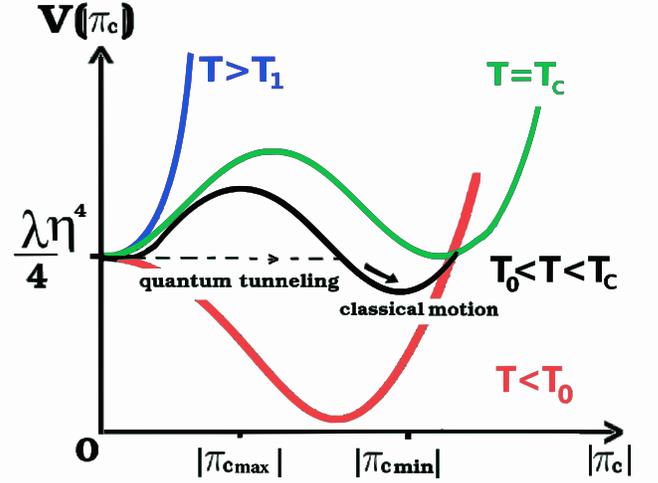}
\caption{Finite temperature effective potential in the core of the string.  
The horizontal dashed line indicates the tunneling of the string from its initial configuration (where the charged fields vanish
everywhere) to the exit point after quantum tunneling.
The arrow toward the minimum indicates the classical decays towards the true vacuum which occurs after quantum tunneling. }
\label{tun}
\end{figure}

This potential (\ref{veff} )is the starting point of our analysis.
There is a 1st order phase transition due to the negative cubic term : this is 
why the potential has metastable vacuum.

As show in Fig.\ref{tun}, for this potential, above certain temperature $T_1$, the vacuum manifold forms a circle in phase space: $|\phi|^2=\eta^2$ and $|\pi_c|=0$.
 As mentioned earlier, such a system admits string solutions: in our case, at temperature below the confinement scale, an embedded neutral string exists.
Above the critical temperature, $T_1$, the potential barrier is infinite in the charged pions direction and no tunneling occurs.
For temperatures between $T_c$ and $T_1$,  $|\pi_c|=0$, is the true VEV  so that the pion string is stable and no skyrmions are generated.

However, for  temperatures between  $T_0$ and the critical temperature $T_c=\frac{\sqrt{6 \la} \eta}{e} \frac{1}{\sqrt{1-\frac{e^4}{3\la \pi^2}}}$,
the system exhibits a 1st order phase transition due to the negative cubic term in the potential.
As a result, the zero charged pion condensate becomes metastable:  $|\pi_c|$ is in the false vacuum and quantum tunneling can occur.
In the following we consider temperatures below $T_c$.

After tunneling  the charged fields take on a non-vanishing expectation value in the true ground state :  $|\pi_c|=v'$ where $v'$is the true VEV.  
The quantum tunneling can happen anywhere on the string, resulting in the nucleation of bubbles at various nucleation points on the string. 
After tunneling in the core of the string, the charged field takes on a non zero value inside each bubble nucleated : $|\pi_c|=v'$ where $v'$ is the VEV
for the potential localized in the core of the neutral string (so that the neutral fields vanish). Since there are two charged fields $|\pi_c|=v'$ describes
a circle of radius $v'$. The bubbles corresponds to instantons  which have a 
non trivial topology in the four scalar fields direction. After tunneling, the 
bubbles expand at almost the speed of light and meet each other to form 
skyrmions, as we will discuss below. 

As a consequence of this tunneling, the core of the neutral string melts and the string generates bubbles at various locations. It is important to note here that we have a global string here with non-zero gradient energy in the winding everywhere (up to
some cut-off distance, say the inter-string separation). Thus the melting
of the core alone cannot break the string. The actual process of the decay
of portions of the string is discussed later in detail in sections
III and IV.

\subsection{Bubble nucleation}

Let us recall some results for bubble nucleation for our specific setting \cite{Karouby:2012kz} . 
The bubble nucleation rate can be roughly estimated to be $ \Gamma  \sim e^{-(B-B_0)}$ where B is the bounce action in euclidean space and $B_0$ is the string background action.

Restricting ourselves to the core of the string, the string background action $B_0$ is assumed to vanish.
First, using the thin wall approximation we obtain the O(4) symmetric action in euclidean spacetime
\ba
S_{sphere}&=\pi^2 \int r^3 d r [ \frac{1}{2} (\frac{\partial {\pi_c}}{\partial_r})^2 + V(\pi_c)] \nonumber \\
           &=-\frac{\pi^2}{2} R^4 \Delta V +2\pi^2 R^3 S_1
\ea
where R is the radius of the bubble in Euclidean space and the potential energy density difference between the two minima is $\Delta V=\frac{e^4}{18 \la} \e (T^2-T_0^2)^2$.

The temperature dependent radius of the nucleated bubble is
 \be R(T)= \frac{3 S_1}{\Delta V}=\sqrt{\frac{3}{2 e}} \frac{1}{\e\sqrt{T^2-T_0^2}} \label{R} \ee
 where the thin-wall parameter $\e$ is a dimensionless quantity that decreases with temperature 
\ba
\label{e2}
\e(T)\hspace{-1mm}=\hspace{-1mm}\sqrt{\frac{e^4}{3 \pi^2 \lambda} \frac{1}{1-\frac{T_0^2}{T^2}}}-1
\ea
 
 The resulting bubble nucleation rate per unit volume is 
\ba
\label{dec}
\frac{\Gamma_{sphere}}{V} \sim P_4 \exp[-\pi^2 \frac{1} {{48\ \la (\sqrt{\frac{e^4}{3 \pi^2 \lambda} \frac{T^2}{T^2-T_0^2}}-1)^3}}]
\ea
where $P_4$, the prefactor has mass dimension 4.

\begin{figure}[h!]
\includegraphics[scale=0.3]{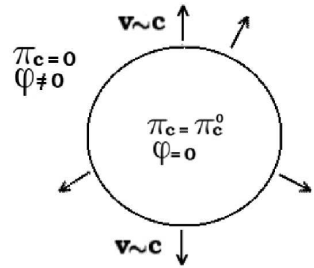} 
\caption{Spherical shape for the instanton configuration\newline
In the idealized case, a sphere of 4-radius R in Euclidean space is generated. The arrows represent the radial expansion of the bubble at almost the speed of light $c$ in Minkowski space. Inside the bubble,
the charged field is in its true vacuum, $\pi_c^0$, and the neutral field vanishes.
} \label{b2}
 \end{figure}

 After tunneling, bubbles of 4-radius $R(T)$ in the Euclidean 
space are nucleated.
 Wick rotating back in Minkowski space-time we see as show in Fig.\ref{b2} that these bubbles will try to expand in Minkowski space.
As the temperature decreases, the string becomes unstable due to the increase
in the bubble nucleation rate. 

We now choose to work in the regime where the decay rate of the metastable string per unit volume is greater than $H^4$
so that the decay of the cosmic string can actually occur in an expanding universe resulting in skyrmion production.

\section{Decay of the string by meson production}

  The manner in which the string decays depends on the rate of
bubble nucleation, and on the ambient temperature. We first
discuss the case with low bubble nucleation rate, when the string 
decays primarily by perturbative
excitations, via meson production. In the next section we then discuss
the case of non-perturbative excitations, when the string decays
into Skyrmions and anti-Skyrmions.

 We have seen above that the U(1) winding of the string can
be trivialized by a tunneling process. This is mediated by an
instanton which converts a localized region on the string
to a topologically trivial mapping on the full vacuum
manifold $S^3$ of the sigma model. It is important to realize 
that the region outside this localized {\it tunneled} region
retains the topologically non-trivial winding on the U(1)
in the $\sigma - \pi_0$ part of the $S^3$. Thus, this instanton
process only destabilizes the core of the string by making it
{\it swell}. This swelling increases as this topologically
trivial region expands. 

Here we note that the metastability of the
string arises from the $-\frac{e^3|\pi_c|^3 }{6\pi}T$ term in the effective potential in Eq.(\ref{veff}). This term 
will become ineffective at low temperatures depending on the
relevant values of gradient energies etc. Thus the chiral field
will start re-orienting away from $\sigma - \pi_0$ plane by
developing non-zero value of $\pi_c$ in regions of strong
gradient energies, e.g. near the core of the string. For sufficiently
low values of the temperature, the chiral field around the whole 
string can re-orient and trivialize the U(1) winding everywhere.
The entire string configuration will then smoothly evolve to the
vacuum configurations with the entire energy of the string converting
into meson excitations. All this will happen without any mediation
of any instanton process.

  However, if the temperature is not low enough, then this process will
remain highly localized near the core of the string, effectively just
leading to thickening of the string. Further decay of the string will
then proceed by instanton process converting a localized region on
the string to roughly uniform orientation of the field. 
For simplicity, we take this
tunneling region to have spherical shape, though our considerations 
below do not specifically require such geometric details.

  In principle, even a single such tunneling region will expand until
the whole string is covered by it and its U(1) winding is 
trivialized completely. For a straight string this will mean
the tunneling region expanding longitudinally along the string
as well as in the transverse direction.  The cutoff scale for
this process will then be provided by typical string-antistring 
separation scale when such expanding regions ({\it swelled cores})
of different string will join with the entire region converted
to topologically trivial mapping. For a string loop, the size of
the loop will provide natural cutoff for this process. 

In the case where
a single tunneling process completes the decay of the string,
the entire energy of the string (which is large as we consider a
global string) will primarily go into meson production which correspond to 
perturbative excitations of the chiral field. However, some nontrivial
structures may arise in between the strings, in a manner described
below. The density of such structures is likely to be 
insignificant in this case.

Now, having explained how mesons arise from the string, we turn to the 
production of baryons through skyrmion production.
In the following we focus on the toplogy of the nucleated instantons
and show that when two bubbles meet, partial or integer skyrmions can form.

\section{String decay by baryon production}

  The string we consider arises here as the 
variation of the chiral field in space is restricted to the
U(1) submanifold of the vacuum manifold $S^3$, spanned by
$\sigma$ and $\pi_0$. The chiral sigma model allows
for genuine topological objects known as Skyrmions corresponding
to non-trivial windings of compactified space $S^3$ to the 
vacuum manifold $S^3$. These Skyrmions are 
identified with baryons as they carry integer baryon numbers
\cite{Skyrme:1961vq,Skyrme:1962vh,bal,Adkins:1983ya}.

  In the last section we discussed the case when the bubble nucleation
rate  by tunneling is low. We argued that in that case the string primarily 
decays into mesons. However, if the decay rate of this tunneling process 
(as given by Eq.(\ref{dec})) is large then a very different mechanism for 
the decay of string may arise.

 In the case of a large bubble nucleation rate, many tunneling regions 
form on the string. These regions then expand and join
with the neighboring regions. After joining, these 
regions keep expanding in the transverse plane as the U(1)
winding remains non-trivial outside such regions. Thus, the whole
process of the decay of the string will still terminate only when
the transverse expansion of these regions is halted by encountering
similar regions from different strings for the straight string case.
For the string loop case the process will stop when similar regions 
from different parts of the same string meet.

More importantly, Skyrmions arise from the intersection region of two neighboring tunneling regions
either on a given string, or on neighboring strings if we consider a string network.

\subsection{Skyrmion production}

\subsubsection{Full Skyrmion in 3 spatial dimension}

  First we consider the simple case where the entire tunneling region 
has some roughly constant value of $\chi$. This is reasonable to expect
as instantons with large field variations in the tunneling region will
have large action and hence will be suppressed. This region will expand and 
meet a neighboring expanding region. If this neighboring region also has 
roughly similar chiral field configuration, with the field pointing 
roughly in the same direction on $S^3$, then these two regions will smoothly 
join and form a larger expanding region. 

\begin{figure}[htbp]
\includegraphics[scale=0.25]{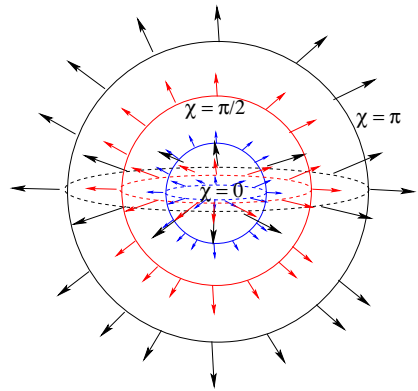}
\caption{A winding number one Skyrmion configuration in 3-dimensions.
Variation of the chiral field is shown on a ball $B^3$. 
Note that $\chi$ varies from 0 (at the center) to $\chi = \pi$ for the 
outer most shell which is the boundary of the ball $B^3$. This leads to 
the compactification of the ball $B^3$ in the physical space to $S^3$}
\label{br31}
\end{figure}

We take a given tunneling region with spherical shape to
correspond to a ball $B^3$ in the physical space. The chiral field
in this region maps  into the vacuum manifold $S^3$.
This entire field configuration can smoothly rotate to a single
point on $S^3$ thereby trivializing the U(1) winding. We first show,
in Fig.\ref{br31}, the topologically non-trivial configuration on a 
ball $B^3$ which corresponds to Skyrmion. The vacuum manifold being
$S^3$, we characterize it by three angles $\theta, \phi$, and $\chi$.
In Fig.\ref{br31}, the polar angle of the arrows (in the physical space) is
taken to represent $\theta$ (on the vacuum manifold). Similarly, the 
azimuthal angle of arrows represents $\phi$. We show, by dotted 
circles, spherical shells, with each spherical shell being 
characterized by the third angle $\chi$ as needed for representing
the three sphere $S^3$. With this  the winding one configuration on
$S^3$, that is a full Skyrmion, will correspond to the variation
of $\chi$ from the value 0 at the center, to $\chi = \pi$, for the 
outermost spherical shell, as shown in Fig.\ref{br31} 
(see ref. \cite{amsskyrm}). Essentially, the ball
$B^3$ represents stereographic projection of the vacuum manifold
$S^3$ with the north pole of $S^3$ being the center of the ball while 
the south pole of $S^3$ being mapped to the boundary of $B^3$ (thereby
compactifying $B^3$ to $S^3$). Here we have represented the winding
one map from (compactified ball) $S^3$ to the vacuum manifold $S^3$
by taking $S^3$ as a suspension of $S^2$ with each
$S^2$ in the suspension being labeled by angle $\chi$ between 0 and 
$\pi$ while each individual $S^2$ is parametrized by angles $\theta$
and $\phi$. 

The three angles $\theta, \phi,\chi$ are related to the
4 scalar fields in the following way : 
\ba
\pi^0= cos\chi 	\ \ \ and\ \ \ \sigma = sin\chi cos\te  \nonumber \\
  \pi^1 = sin\chi sin\te cos\phi	\ \ \ and \  \ \ \pi^2= sin\chi sin\te sin\phi
\ea

where the charged fields $\pi^\pm$ are defined in terms of $\pi^1$ and $\pi^2$  in Eq.\ref{pi}.

\subsubsection{Full Skyrmion in 2 spatial dimensions}

Consider now the situation when the neighboring region points in roughly 
opposite direction, forming small patches in the respective regions on $S^3$.
As the situation with $S^3$ is difficult to visualize, we will consider the 
simpler situation of two-dimensional Skyrmion, with physical space
being two dimensional and tunneling region being a disk. The vacuum
manifold in this case will be $S^2$ and a winding one Skyrmion will
correspond to the stereographic projection of the vacuum manifold $S^2$
on the disk in the physical space. This is shown in Fig.4 where the
south pole of the vacuum manifold $S^2$ is mapped to the center of the
disk while the north pole is mapped to the boundary of the disk, thereby
compactifying the disk to $S^2$.

\begin{figure}[htbp]
\includegraphics[scale=0.5]{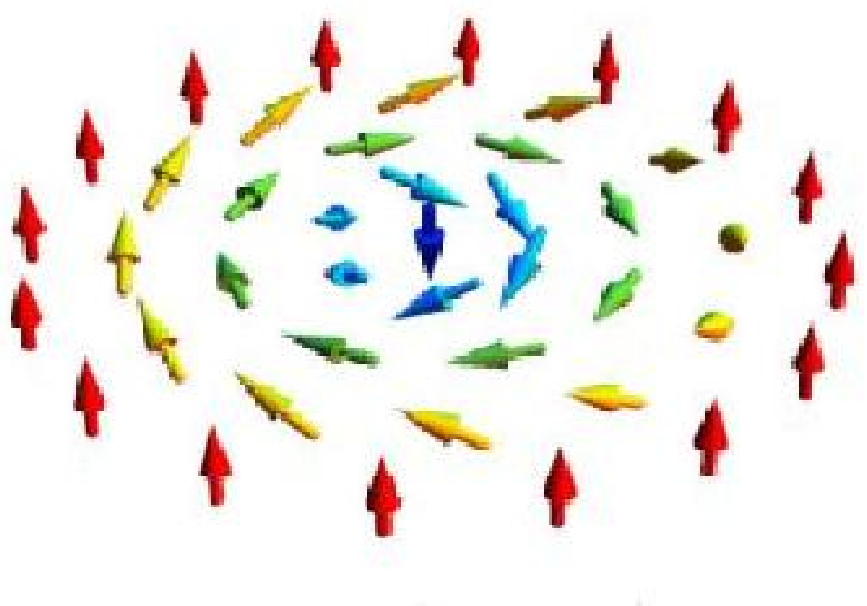}
\caption{A winding number one Skyrmion configuration in 2-dimensions. 
Variation of the field is shown on a disk.
South pole of the vacuum manifold is mapped to the center of the disk
while the north pole is mapped to the boundary of the disk compactifying 
it to a two-sphere $S^2$.}
\label{br4}
\end{figure}

\subsubsection{Instanton meeting with opposite patches in the vacuum manifold}

  We now consider the situation of the string with two neighboring
tunneling regions corresponding to roughly opposite patches on the vacuum
manifold. We note that it should be expected that the tunneling region will 
form small patches on $S^3$  as presumably only such instantons 
will dominate the tunneling process. The size of these patches are of
crucial importance here, and these will depend on the calculations of
the determinant of fluctuations around the semi-classical instanton
background. If the fluctuations are strong, then we expect the patches
to have significant size. The size of the patch will depend on
the temperature and one will expect that for high temperature
tunneling the patch can be large. We will explain the underlying physics
here by using two-dimensional example and then generalize the arguments
for the $S^3$ case. 

\begin{figure}[h]
\includegraphics[scale=0.25]{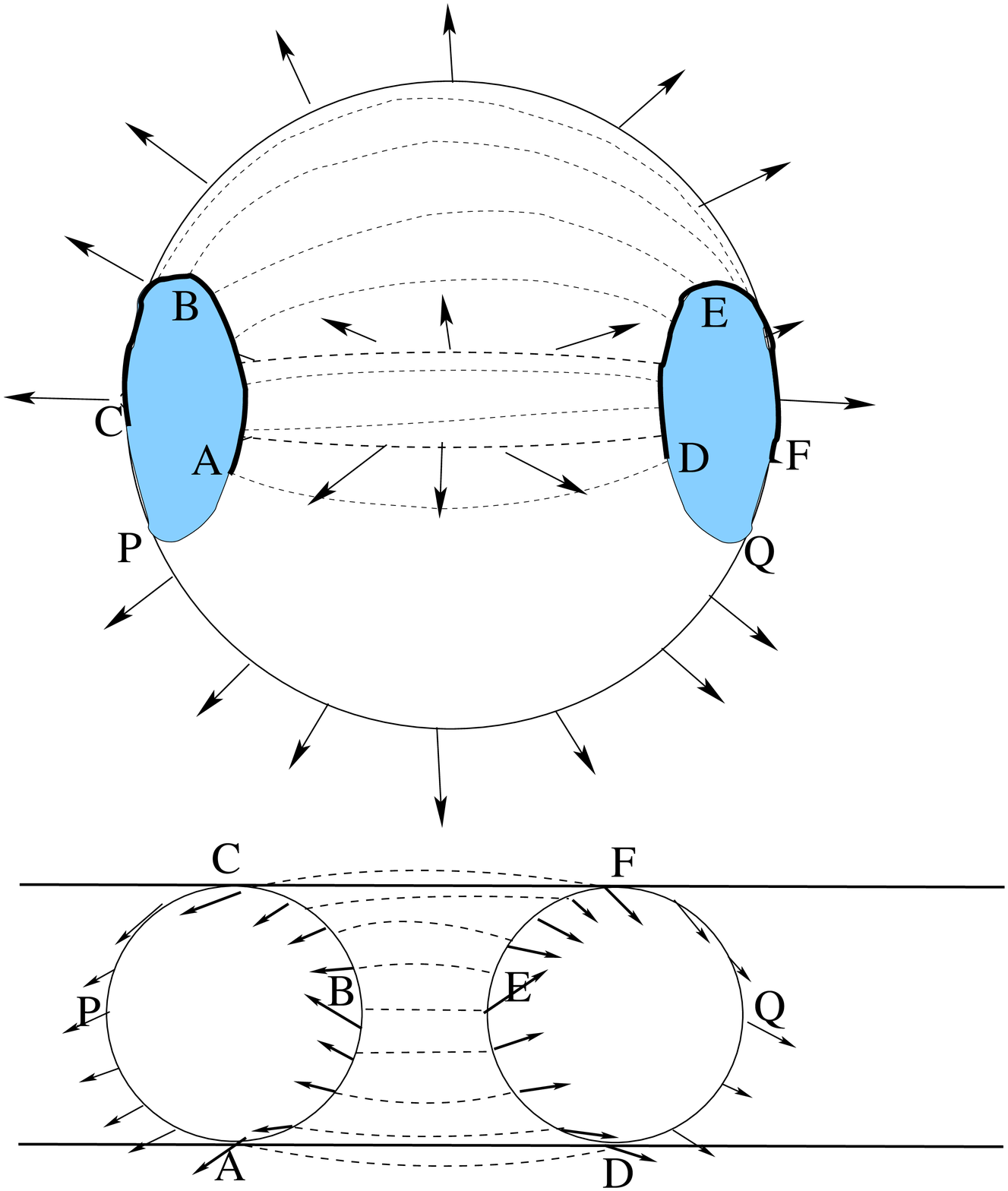}
\caption{Meeting of two instanton (bubble) configurations for a 
two-dimensional case. 
The figure on top shows the two patches (blue
color) ABCP and DEFQ on the vacuum manifold $S^2$ representing
the field configuration in two neighboring regions inside the string.
In the bottom figure, the string is represented as a strip.
 The region in-between the two 
patches is only sensitive to the parts of the boundaries of the two patches, 
marked as ABC for the left patch and DEF for the right patch. The field
configuration for the intermediate region, therefore, is seen to correspond
to the dashed lines shown on the $S^2$ in the top figure.}
\label{br5}
\end{figure}

The top figure in Fig.\ref{br5} shows the two patches (blue
color) ABCP and DEFQ on the vacuum manifold $S^2$ representing
the field configuration in two neighboring regions inside the string.
The string here is shown as a strip (lower figure in Fig.\ref{br5}) as appropriate for
the two-dimensional case. An important thing to note is the field
configuration in-between the two patches in this strip. 
As the two regions meet, the chiral field in the junction region will
evolve to smoothly interpolate between the two corresponding values on
the vacuum manifold $S^2$. This is essentially the statement of the so called 
{\it geodesic rule} in the theory of topological defect formation via 
domain coalescence, known as the Kibble mechanism \cite{kbl}. Different
tunneling regions in a given segment of string form the uncorrelated 
domains of the Kibble mechanism here. However, the region in-between the two 
patches is only sensitive to the parts of the boundaries of the two patches, 
marked as ABC for the left patch and DEF for the right patch. The field
configuration for the intermediate region is then seen to correspond
to the dashed lines shown on the $S^2$ in the top figure in Fig.\ref{br5}.

   The conclusion then is that if the two neighboring tunneling regions
in the string core correspond to the two opposite patches as shown in
Fig.\ref{br5}, then the entire region consisting of the two patches, along with
the region in between, will correspond to covering of $S^2$ by a surface
spanned by the dashed curves on $S^2$ shown in Fig.\ref{br5}, 
including the regions
of the two  patches. To understand the evolution of this field configuration,
we recall the discussion of formation of Skyrmions via Kibble mechanism
in phase transitions from ref. \cite{amsskyrm}. It was argued in \cite{amsskyrm}
that, in contrast to the formation of domain walls, strings, and monopoles
in phase transitions, Skyrmions never form with full integer winding
number in a phase transition. This is simply due to the fact the 
integer winding Skyrmion corresponds to a mapping from {\it compactified}
three dimensional space to the vacuum manifold $S^3$. Compactification
of three dimensional space requires constant boundary conditions which
is virtually impossible to achieve in a phase transition where domains
form with randomly varying field configurations. However, the so called
{\it partial winding} Skyrmions will routinely form. These correspond
to field configurations which cover part of $S^3$. It was argued in 
\cite{amsskyrm} that if a partial winding configuration covers more than
50 \% of the vacuum manifold $S^3$, then it will evolve to a complete
integer winding Skyrmion. If the configuration covers less then 
50 \% of $S^3$ then it will shrink down and should decay in 
perturbative meson excitations. This is also seen in simulations of
the formation of textures in the context of early Universe \cite{turok}. 

\section{Probability of skyrmion formation}

 Applying the same arguments to our present case, we see that
if the two patches do not lie reasonably opposite to each other on
$S^2$ the combined region of the two neighboring patches will
cover less than half of the vacuum manifold. In such a situation, 
the whole configuration will shrink down on $S^2$ decaying into mesons. 

However, for the case shown in Fig.\ref{br5} when the two patches are opposite
to each other on $S^2$, the field in this entire region is likely to
cover more than half of  the $S^2$. Further evolution of such a 
configuration should lead to integer winding Skyrmion.
It is simple to see that if a junction of two neighboring domains
leads to formation of Skyrmions (anti-Skyrmions), then the next 
non-trivial intersection is likely to lead to the formation of anti-Skyrmions
(Skyrmions). If the next intersection happens to lead to topologically 
trivial configuration, that simply leads to the deformation of the Skyrmion.

  The probability of having the two neighboring patches in the string which correspond to opposite patches on $S^2$ can be estimated as follows : 
The average size of the patch on $S^2$ is 1/8 of the surface area of $S^2$
(see, ref.\cite{amsskyrm}). With 8 average patches on $S^2$, the probability
of the second patch being opposite to the first one will be 1/8. Thus
we will conclude that the probability of the junction region of two 
tunneling regions to evolve into full winding Skyrmion is 1/8. This
is a crude estimate assuming large patches (which may be reasonable for
high temperature tunneling). If the patches are small then this probability
will be suppressed. 

  The extension to three-dimensional case is then straightforward.
We go back to Fig.\ref{br31} showing the configuration of $S^3$ Skyrmion. The
two neighboring regions will be taken to be opposite patches on
this $S^3$ (like top figure in Fig.\ref{br5}). The string will now be a
tube (instead of a strip as in Fig.\ref{br5}) with the two neighboring regions
forming two balls inside this tube. The dashed curves of Fig.\ref{br5} will
now connect the boundaries of the two balls and will lead to a covering
of part of $S^3$. As we have argued for the two-dimensional case,
here also, if the two patches are opposite on $S^3$ then it will be
expected to cover more than half of $S^3$ (using geodesic rule). The
evolution of this field configuration should then lead to a full
winding Skyrmion.

 The probability of having two opposite patches
is then 1/16 by extension of the above argument (as average volume
of a patch on $S^3$ is 1/16 of the total volume \cite{amsskyrm}). 
This is similar to the probability of 0.05 for the formation of partial 
winding Skyrmions with more than 50 \% cover of $S^3$ as estimated 
in \cite{amsskyrm} (and consistent with the results found in texture 
simulations \cite{turok}). 

We also mention that we are considering 
the chiral limit here with massless pions. With finite pion mass the 
sigma model has an explicit symmetry breaking term. This makes one point
on the vacuum manifold (say south pole) as the true vacuum, while the 
opposite point, the north pole, has the highest potential energy. It 
was shown in \cite{kpst} that with such explicit symmetry breaking, the 
production of  Skyrmions-anti-Skyrmions may be significantly enhanced. 
The same behaviour is expected in our case. For example, even if the field 
configuration in the two regions in Fig.\ref{br5} covers less than
half of the vacuum manifold, if it overlaps with the north pole
(the point with highest potential energy) then the entire field 
configuration will start rolling down towards the true vacuum, thereby
covering the entire vacuum manifold.

 Even with weak explicit symmetry
breaking such effects will be present. For strong explicit symmetry
breaking one can even think of the possibility that a single instanton
having a patch around the north pole may evolve into a full winding
Skyrmion as the field rolls down everywhere towards the true vacuum
thereby covering the entire vacuum manifold.  

Finally, we note that so far, we considered the encounter of the two bubble regions happens only in the 
rest frame of the string. It is important to 
remember that cosmic strings are actually moving at a speed close to the speed of light.
This means that Skyrmions will actually spread outside the string. 
Since the Skyrmions anti-Skyrmions pairs that are formed can be interpreted as a 
production of baryon-anti baryons, the string can be seen as a 
baryon-antibaryon generator. 

\subsection{Estimates of baryon production by string decay}

The number of instanton formed per unit volume per unit time is equal to 
the decay rate $\Gamma$. We are considering the case when the decay
rate is large compared to the Hubble scale, thus we will assume that
within a Hubble time, a certain number of bubbles per unit length,
$n(t)$, nucleate on the string with

\begin{equation}
n(t) =  \pi R_0^2 \Gamma H(t)^{-1}
\end{equation}

Here  $R_0= \frac{1}{\sqrt{\la} \eta}$ is the estimated radius of the 
neutral string. For simplicity, we will take all these bubbles to be 
of same size. Then, the number per unit length of two (neighboring) 
domain intersections is also equal to $n$. With the probability of
an integer winding Skyrmion/anti-Skyrmion (i.e. a baryon/anti-baryon)
formation per junction being 1/16, we get the number of baryon-anti-baryon
produced per unit length $n_b$ from the string as, $n_b = {n \over 16}$.

The largest number of bubbles
 nucleated on the string during Hubble time can be simply estimated
\begin{equation}
n \simeq (2R(T))^{-1} \simeq \sqrt{\frac{ e}{6}} \e(T)\sqrt{T^2-T_0^2} \label{n2}
\end{equation}
 where $R(T)$given by Eq. (\ref{R}) is the radius of the nucleated bubble
 and $\e(T)$ in Eq. (\ref{e2}) is the thin wall parameter.
For each value of the temperature
 $T$, we can thus determine the value of $n$, and hence $n_b$.

The net baryon density produced by these strings will depend on
the string density. As the relevant temperature regime is not expected
to be too much below the transition temperature here, it is unlikely
that the string will enter into any scaling regime. The estimate of
string density then becomes difficult as it will depend on string
velocity, friction coefficient of string moving through plasma, and
the shrinking rate of the string (without decay as discussed here).

\subsection{Baryon production by a dense string network}

  Another  different and straightforward estimate for the formation of 
Skyrmions-anti-Skyrmions can be obtained by direct adaptation of 
calculations of refs.\cite{amsskyrm,turok}.
Let us consider the situation of a dense network of pionic strings
with a high rate of instanton mediated decay process. With all the strings
essentially decaying into multiple tunneling regions, the entire physical
region eventually converts to a multi-domain region, almost
the same as resulting after a phase transition with domains corresponding
to correlation domains. The typical domain size here is a combination
of two length scales : the typical tunneling region size when it
intersects another such region, and the average inter-string separation.
Assuming these two scales are similar for simplicity, 
the situation reduces to that of a phase transition as discussed
in refs.\cite{amsskyrm,turok}. Here one even does not need to assume any spread
in the value of the order parameter field in a given domain. The estimate of 
Skyrmion-anti-Skyrmion formation in this case is simply 0.05 per domain.    
Here also, we mention that the estimate of 0.05 Skyrmions per domain
was for the case of chiral limit. With explicit symmetry breaking we
expect enhanced production of Skyrmions.

An important difference in this formation and the earlier one discussed
above (with two domains) is that in the earlier case net Skyrmion number
was zero as required by baryon conservation in QCD. This is because
every junction corresponding to a given winding was followed by the next
non-trivial junction with opposite winding. This correlation is not
present in the second picture of Skyrmion formation. In this second
case, the conservation of baryon number is not explicitly present.
An attempt to incorporate baryon number conservation in this type
of picture has been discussed in ref.\cite{sunil}. An interesting aspect is that
 a baryon asymmetry can also be generated.

\section{Conclusion}

In this work we explained and quantified a possible mechanism to create
baryons from pion-like cosmic strings namely skyrmion-anti skyrmion 
production by a metastable string.
We showed that the embedded string of the linear sigma model (which can be metastable provided we choose the right parameter for the coupling constant)
generates skyrmions with partial or integer winding number.
We quantified the integer winding number skyrmions or anti-skyrmions nucleation per unit length per Hubble time 
which can be interpreted as baryon anti-baryons production. Decay of the string can also arise through meson excitations.
 Finally we commented on the baryon production from a dense string 
network and estimated the probability of skyrmion-anti-skyrmion formation 
to be about 5 \% per domain.
 
There are many effects to consider such that the effect of the string
velocity, friction coefficient of string moving through plasma, and
the shrinking rate of the string (without decay as discussed here). 
We hope to discuss these aspects in a future work.

\begin{acknowledgments}
We would like to thank Robert Brandenberger.
Johanna Karouby is supported in part by the U.S. Department of Energy under cooperative research agreement DE-FG02-05ER-41360 and by an NSERC PDF fellowship.\end{acknowledgments}

\end{document}